\documentclass[
amsmath,
amssymb,
aps,
twocolumn,
superscriptaddress]{revtex4-2}

\usepackage{graphicx}
\usepackage{dcolumn}
\usepackage{bm}
\usepackage{hyperref}
\usepackage{float}

\begin{document}

\title{Bridging deep learning force fields and electronic structures with a physics-informed approach}

\author{Yubo Qi}
\affiliation{Department of Physics, Northeastern University, MA 02115, USA}
\affiliation{Department of Physics, University of Alabama at Birmingham, AL 35233, USA}

\author{Weiyi Gong}
\affiliation{Department of Physics, Northeastern University, MA 02115, USA}

\author{Qimin Yan}
\email{q.yan@northeastern.edu}
\affiliation{Department of Physics, Northeastern University, MA 02115, USA}

\begin{abstract}
This work presents a physics-informed neural network approach bridging deep-learning force field and electronic structure simulations, illustrated through twisted two-dimensional large-scale material systems. 
The deep potential molecular dynamics model is adopted as the backbone, and electronic structure simulation is integrated. 
Using Wannier functions as the basis, we categorize Wannier Hamiltonian elements based on physical principles to incorporate diverse information from a deep-learning force field model.
This information-sharing mechanism streamlines the architecture of our dual-functional model, enhancing its efficiency and effectiveness.
This Wannier-based dual-functional model for simulating electronic band and structural relaxation (WANDER) serves as a powerful tool to explore large-scale systems.
By endowing a well-developed machine-learning force field with electronic structure simulation capabilities, the study marks a significant advancement in developing multimodal machine-learning-based computational methods that can achieve multiple functionalities traditionally exclusive to first-principles calculations.
Moreover, utilizing Wannier functions as the basis lays the groundwork for predicting more physical quantities.
\end{abstract}


\maketitle

\section{Introduction}
First-principles calculations based on density functional theory (DFT) have emerged as a powerful tool in predicting and calculating physical properties~\cite{Kohn96p12974}. 
Their accurate descriptions of structural parameters, energy landscapes, and electronic structures, directly comparable with experimental measurements, have yielded deep physical insights~\cite{Neugebauer13p438}.
However, as the system gets large, the computational cost of DFT calculation increases dramatically, making the simulations of many scientifically important solid-state systems, such as twisted 2D materials, heterostructures, and crystals with defects, technically challenging and even impossible. 
The recent developments in machine learning techniques demonstrate promising potential in addressing this challenge~\cite{Carleo19p045002,Schmidt19p83,Behler21p10037,Unke21p10142,Zhang18p143001,Artrith11p153101,Sosso12p174103,Deringer19p1902765,Chen22p718,Shapeev16p1153,Thompson15p316,Behler07p146401,Bartok10p136403,Unke19p3678,Schutt18p}. 
Specifically, machine learning force fields have been shown to produce highly accurate atomic forces and crystalline energies~\cite{Zhang18p143001,Artrith11p153101,Sosso12p174103,Deringer19p1902765,Chen22p718,Shapeev16p1153,Thompson15p316,Behler07p146401,Bartok10p136403,Unke19p3678,Schutt18p,Choudhary23p346,Park21p73,Haghighatlari22p333,Batzner22p2453,Husic20p194101,Xiao23p7027}. 
In these force fields, the local atomic environment is digitized, and neural networks are employed to identify the relationship between the local atomic environment and atomic forces. 
Despite being a powerful tool for simulating atomic forces and optimizing structures, machine learning techniques also demonstrate promising potential in electronic structure simulation.
Recently, an approach based on a message-passing neural network (referred to as DeepH) has been developed to simulate the first-principles Hamiltonian and subsequently acquired the electronic structures~\cite{Li22p367,Li23p321,Gong23p2848}.
Each of these works reproduces a primary functionality of DFT calculations with a significantly increased efficiency, representing a substantial step in developing machine learning-based computational methodologies.

According to the ``nearsightedness'' model proposed by Prodan and Kohn, the change of potential at a distant position has little effect on local electronic properties~\cite{Prodan05p11635}. 
This principle implies that both atomic forces and electronic structures are local physical properties and should be able to be predicted by similar models. 
However, contrary to the remarkable progress in machine-learning force fields~\cite{Zhang18p143001,Wang18p178,Xiao23p7027,Wu23p144102,Wu21p024108,Choudhary23p346,Park21p73,Xu20p16278,Ouyang23p20890}, there remains a notable scarcity of deep-learning models for solid-state electronic structure simulation despite the success of DeepH~\cite{Li22p367}.
This observation prompts the question of whether it is feasible to bridge this gap by developing a method endowing existing machine learning force fields with the capability for electronic structure simulation.
If so, the inquiry extends to identifying the guidelines for designing such a dual-functional model that not only ensures optimal efficiency and transferability but also attains high levels of accuracy.

Inspired by these challenges, we develop a physics-informed neural network approach for simulating both atomic and electronic structures. 
This approach uses the Wannier functions (generated from atomic orbitals) as the basis, and adopts the framework of a well-established and widely adopted machine learning force field model~\cite{Zhang18p143001,Wang18p178}.
Information learned from the force field model is used to facilitate the simulation of electronic band structures, 
and this information-sharing mechanism streamlines the architecture of our dual-functional model, enhancing its efficiency and effectiveness.
To demonstrate the performance of this Wannier-based dual functional model for simulating electronic band and structural relaxation (WANDER), we take twisted MoS$_2$ bi-layer systems as illustrative examples since the Moir\'e structures of 2D materials host a wide range of novel physical properties~\cite{Carr20p748,Andrei20p1265,Bistritzer11p12233}. 
This work aims to bridge deep learning-based models for atomic-structure and electronic-structure simulations and provide insights on developing machine-learning-based computational methods offering multiple functionalities of first-principles calculations. 
Moreover, this research presents a feasible approach to augment machine learning models for electronic band structure simulations, potentially bridging the gap between the scarcity of machine learning-based electronic structure simulation approaches and the prosperity of machine learning-based force fields.

This work adopts Wannier functions as the basis, which are a complete set of real-space orthogonal functions acquired from the Fourier transforms of the Bloch functions~\cite{Kohn59p809}. 
Wannier functions have been widely adopted in the research of theoretical solid-state physics since they capture the essential physics of a material’s electronic structure~\cite{Souza01p035109,Marzari12p1419}. 
For example, the $k$-space Hamiltonian matrix elements, whose eigenvalues yield the electronic band structure, can be acquired through Wannier interpolation as
\begin{multline}
     H(\mathbf{k})_{m\alpha, n\beta} =   \\
     \sum_{m,n}\sum_{\alpha,\beta}\sum_{\mathbf{a}}
     e^{i\mathbf{k}\cdot\mathbf{a}}
     \left<w_{m\alpha}\left(\mathbf{r}\right)
     \left|H\left(\mathbf{r}\right)\right|
     w_{n\beta}\left(\mathbf{r+a}\right)\right>.
\end{multline}
Here, $H\left(\mathbf{k}\right)$ and $H\left(\mathbf{r}\right)$ are the Hamiltonians in the reciprocal space and real space, respectively. $\mathbf{a}$, $m\left(n\right)$, and $\alpha\left(\beta\right)$ are the indices of cells, atoms, and Wannier functions, respectively. 
In the WANDER, the Wannier Hamiltonian elements, which determine the electronic band structures, are divided into three categories (on-site interactions, intra-layer hopping integrals, and inter-layer hopping integrals) as suggested in a previous tight-binding model~\cite{Wang22p159}. 
Our study shows that the input representation of each category should take different information from the deep learning force field model for improved performance in terms of accuracy, efficiency, and transferability.
The selection of these representations is guided by the underlying physical mechanisms, aligning with the concept of physics-informed machine learning~\cite{Karniadakis21p422}. 
While our focus is primarily on atomic and electronic structure simulation, utilizing Wannier functions as the basis lays the groundwork for integrating additional functionalities into our approach. 
This expansion broadens the predictive capabilities of our model to encompass a diverse array of physical quantities calculable based on Wannier functions, including but not limited to spin Hall conductivity~\cite{Wang06p195118,Zhou19p060408}, shift current~\cite{Ibanez18p245143}, transport properties~\cite{Pizzi14p422}, and electron-phonon coupling~\cite{Giustino07p165108, Ponce16p116}.

\section{Results}
\subsection{Overview of Wannier functions}
Wannier functions for a specific system are non-unique since a Bloch function ${\phi_{n\mathbf{k}}}$ can adopt a phase $e^{i\phi\left(\mathbf{k}\right)}$ as an arbitrary function of $\mathbf{k}$. 
There are two widely adopted approaches for acquiring well-defined Wannier functions. The first approach is to minimize their spreads (through a process known as localization) to acquire maximally localized Wannier functions (MLWFs)~\cite{Souza01p035109,Marzari12p1419}. 
MLWFs are exponentially decaying and widely adopted as the basis for constructing first-principles, tight-binding models~\cite{Koshino18p031087,Garrity21p106}. 
However, MLWFs change with structural distortions dramatically, whose trend is challenging to track; for Chern insulators, it is even infeasible to construct exponentially localized Wannier functions~\cite{Brouder07p046402}. 
An alternative approach, which can get rid of these challenges, is to use Wannierized atomic orbitals, without any localization steps, as the basis~\cite{Zhang10p065013,Shi21p381,Ponraj16p462001}. Wannier functions acquired through this method have larger spreads, leading to increased computational costs. 

In this work, we adopt Wannier functions generated from atomic orbitals with finite localization as the basis, and the computational procedure is as follows. 
First, we calculate the MLWFs of the ground state MoS$_2$, which is a trivial insulator. Then, we employ atomic orbitals to approximate the MLWFs. 
These orbitals should be linearly independent and form a complete set (see Supplementary Information Section 1 for details). 
Finally, we use the atomic orbitals as the initial guess for projection, acquiring Wannier functions and minimizing their spreads for finite (40 in this study) iterations with the Wannier90 package~\cite{Mostofi14p2309,Pizzi20p165902}. 
The acquired ``semi-localized'' Wannier functions are used as the basis. 

The essence of this approach closely resembles that of using Wannierized atomic orbitals as the basis. 
Using the atomic orbitals, which approximate the MLWFs of the ground state, as the initial projection will not introduce any error as long as the atomic orbitals form a complete set.
However, the subsequent localization iterations may introduce errors. 
Here, we take the assumption that the semi-localized Wannier functions exhibit slight variation across different structures. 
In other words, the localization process for finite iterations decreases the spread of Wannier functions with slight shape changes (see Supplementary Information Section 2 for details). 
Our results, based on bi-layer MoS$_2$, align with this assumption, demonstrating that using ``semi-localized" Wannier functions as the basis has little impact on the model's accuracy. 

It is worth mentioning that localization with finite steps is completely optional and can be viewed as a trade-off between accuracy and efficiency. 
While it may introduce errors, the localization process also breaks the symmetry of the Wannier functions, making this approach unsuitable for investigating topological properties. 
To overcome these challenges, future research may focus on developing symmetry-adapted Wannier functions or designing standardized Wannier functions with small spreads~\cite{Sakuma13p235109,Lei20p187402}. 
However, these topics are beyond the scope of this study and will be pursued in future research.

\subsection{Neural network scheme}
\begin{figure*}
\centering
\includegraphics[width=16.0cm]{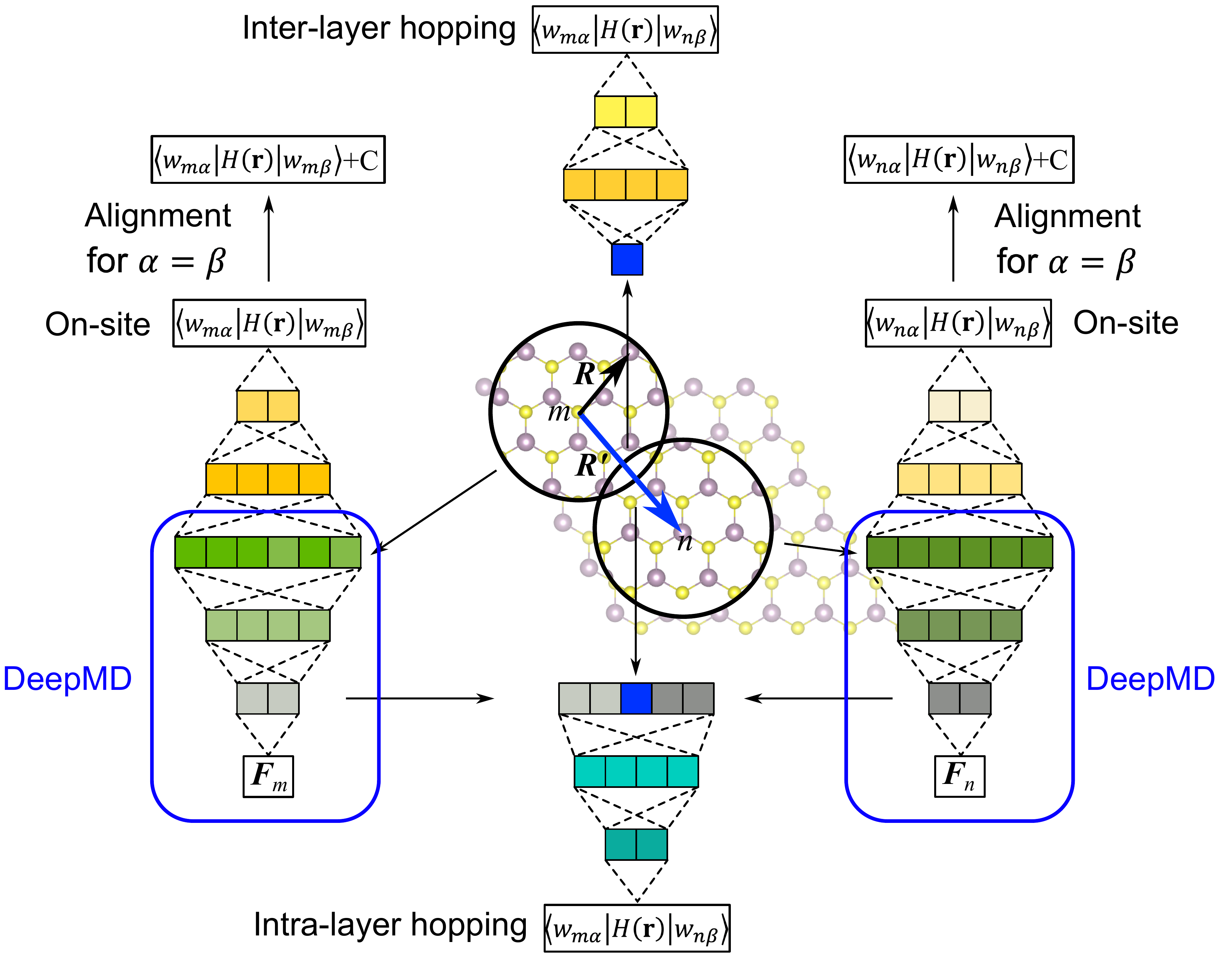}
\caption{\textbf{Schematic illustration of the architecture of the WANDER.} First, an atomic force model is trained with the DPMD method, as indicated by the blue rectangles. 
The last hidden layer, referred to as RAER in this work, contains information about the local atomic environment. 
On-site interaction, intra-layer hopping integrals, and inter-layer hopping integrals are predicted with inputs for DPMD, RAERs plus relative-position vectors, and relative-position vectors, respectively.  
}
\label{f1}
\end{figure*}

The schematic plot of the architecture of WANDER is shown in Fig.~\ref{f1}.
The first step is to train a force field model, as indicated by the blue rectangles in Fig.~\ref{f1}. 
Here, we adopt the deep potential molecular dynamics (DPMD) model introduced in Ref.~\cite{Zhang18p143001,Wang18p178}. 
The local environment of a specific atom is transformed into input data by stacking the 4-dimensional vector $\left(\frac{1}{R},\frac{R_x}{R^2},\frac{R_y}{R^2},\frac{R_z}{R^2}\right)$ contributed by each neighboring atom inside the cutoff radius $R_C$. 
Here, $R$ is the length of the vector $R$ pointing from the central atom to its neighbor. $R_x$, $R_y$, and $R_z$ are the three components of $\bm{R}$. 
The atomic forces in Cartesian coordinates $\left(f_x,f_y,f_z\right)$ serve as the output. Two hidden layers, with 32 and 16 nodes, respectively, connect the input and output, forming a feed-forward network. 
Between the input and different hidden layers, data are transferred through a linear transformation $\bm{d}_{k+1}=\bm{W}_k\bm{d}_k+\bm{b}_k$ followed by a rectified linear unit (ReLU) activation function. 
Here, $\bm{W}_k$ and $\bm{b}_k$ are learnable weights and biases at the $k$th step. 
Between the last hidden layer and the output layer, only a linear transformation is applied. 

The Wannier Hamiltonian elements are classified into three categories as on-site interaction, intra-layer hopping integrals, and inter-layer hopping integrals. 
This classification is in accordance with the expression of the tight-binding Hamiltonian in Ref~\cite{Wang22p159}
\begin{equation}
H_{\rm{TB}}=-\sum_{\left<m,n\right>}t_{mn}^{\rho=0}c_m^{\dag}c_n
-\sum_{m,n}t_{mn}^{\rho=1}c_m^{\dag}c_n
+\sum_{m}{\epsilon}c_m^{\dag}c_m,
\end{equation}
where $\rho=0$ and $\rho=1$ stand for atoms $i$ and $j$ located at the same and different layers, respectively, and $t_{mn}$ and $\epsilon$ are on-site and hopping integrals. 

\begin{figure*}
\centering
\includegraphics[width=17.0cm]{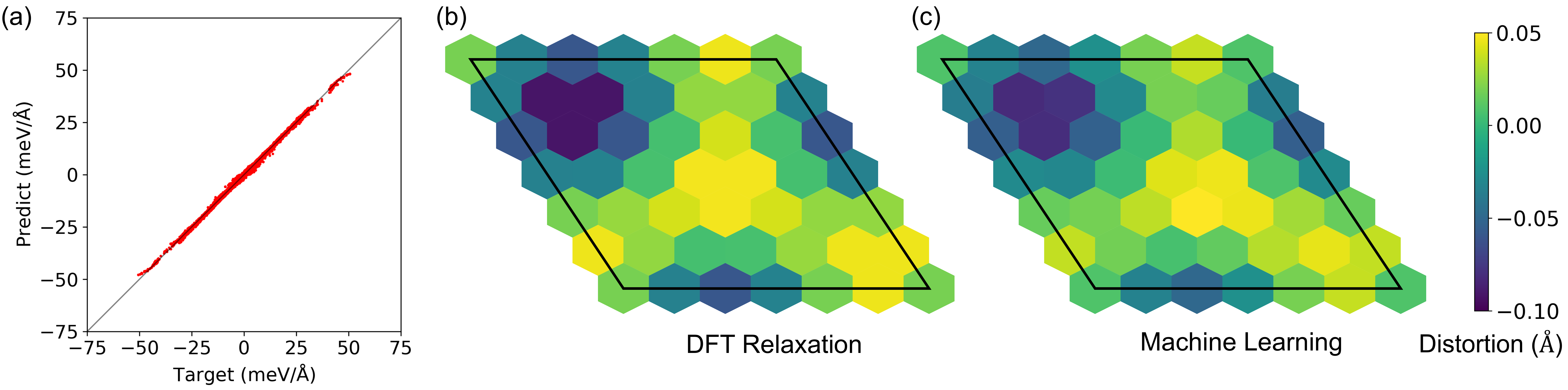}
\caption{\textbf{Performance of the WANDER for atomic-force predictions.} \textbf{a,} The parity plot for the atomic forces. 
\textbf{b,c,} Out-of-plane distortions induced by the Moir\'e potential given by the \textbf{(b)} DFT calculation and \textbf{(c)} machine-learning model prediction. 
For a clear view, we only plot the distortions in the bottom layer.
}
\label{f2}
\end{figure*}

The three categories of Wannier Hamiltonian elements (on-site interaction, intra-layer hopping integrals, and inter-layer hopping integrals) have different underlying physical mechanisms and should be predicted with different models, as shown in Fig.~\ref{f1}. 
Similar to an atomic force, an on-site interaction is a single-body term that is determined by the local atomic environment; predictions for atomic forces and on-site interactions share the same input. 
Hopping integrals are two-body interactions, and the relative positions of the two atoms should be incorporated into the input. 
The relative position is represented by a 4-dimensional vector $\left(\frac{1}{R^\prime}, \frac{{R^\prime}_x}{R^{\prime2}}, \frac{{R^\prime}_y}{R^{\prime2}}, \frac{{R^\prime}_z}{R^{\prime2}}\right)$, 
where $R^\prime$ is the length of the vector $R^\prime$ between the two atoms. 
To predict intra-layer hopping integrals, we extract the 16-dimensional vector of each atom from the last hidden layer in the DPMD model and stack them with the relative position vector as the input. 
Here, the last hidden layer in DPMD contains essential information about the local atomic environment and thus can be viewed as a reduced atomic environmental representation (RAER).
Later, we will show that this input containing RAER performs better than stacking the DPMD first-layer (input for DPMD) vector and the relative-position vector. 
The input for predicting the inter-layer hopping integral is the 4-dimensional relative-position vector. 
This is because there is no atom in the inter-layer gap affecting the Hamiltonian, making the hopping integral mainly depend on the relative positions of the two atoms. 
The deep learning architecture for predicting Wannier Hamiltonian elements is the same as that of DPMD, which comprises two hidden layers.

\subsection{Dataset for training and vacuum level alignment}
To train the model for a force field, we build a database with 10690 different configurations of 4$\times$4 bi-layer MoS$_2$. 
In each structure, one of the MoS$_2$ layers is parallelly displaced with an arbitrary 3-dimensional vector. 
As a result, these structures have different inter-layer translations or inter-layer gaps. 
Perturbations on atomic positions are applied to 90\% of the configurations to sample a larger group of structures (see Supplementary Information Section 3 for details). 
We carry out DFT calculations on all the configurations and acquire the atomic forces. 
80\% of the data are used for training, and the remaining 20\% are used for testing.  

To train models for predicting Wannier Hamiltonian elements, we employ a subset of 662 configurations from the dataset containing 10,690 configurations for atomic-force prediction. 662 configurations give approximately81,500,000 non-zero (with absolute magnitude larger than 1 meV) Wannier Hamiltonian elements. This quantity is substantial enough to facilitate the learning of the underlying rules.

To validate our model, we compare the DFT results of twisted bilayer MoS$_2$ structures. 
The twisted structures are undoubtedly not present in the training set composed of parallel non-twisted bilayer structures, serving as an outstanding demonstration of the transferability of this model. 
In twisted structures, one of the MoS$_2$ layers should be strained to be commensurate with periodic boundary conditions. 
For a large Moir\'e superlattice, the numbers of atoms in different layers can be different. In this case, a correction term should be added to the diagonal elements $\left< w_{m\alpha}\left| H\right| w_{m\alpha}\right>$ in the Wannier Hamiltonian to align the vacuum levels of the two layers. 

In DFT calculations, the energy at the vacuum level $E_{vl}$ is considered as zero energy~\cite{Hinuma14p155405}. 
$E_{vl}$ is determined by the Hamiltonian $H_{\rm{DFT}}$, which depends on the number of atoms in a supercell~\cite{Choe18p196802,Zhang16p015038}. 
The removal or addition of atoms in a supercell invariably impacts the vacuum level of the entire system. 
In this model, the Wannier Hamiltonian elements are learned based on local atomic environments in those structures with equal numbers of atoms in each layer. 
To apply this mode to twisted bilayer systems, the vacuum level of the two layers with different numbers of atoms should be aligned by adding a constant $C$ to the predicted diagonal terms $\left<w_{m\alpha}\left|H\right|w_{m\alpha}\right>$, as shown in Fig.~\ref{f1}. 
Our calculations show that $C$ for a layer depends on the strain $\varepsilon$ and the ratio between the number of atoms $N_{\rm{out}}$ outside of the layer and the number of atoms $N_{\rm{in}}$ inside the layer. 
We build a shallow-level machine learning model (polynomial regression) to predict $C$ with $\varepsilon$ and $N_{\rm{out}}/N_{\rm{in}}$ as the input. 
The model performs quite well with an R-squared value of 0.9999999 (see Supplementary Information Section 4 for details). 

\subsection{Model performance for atomic forces prediction}

Fig.~\ref{f2} (a) shows the parity plot for atomic force predictions. 
For the testing dataset, the model predictions match DFT results well with a 0.2 meV/\AA\ mean absolute error (MAE). 
To demonstrate the transferability of this model, we consider a bilayer MoS$_2$ structure with 8.95$^{\circ}$ twisting angle and 201 atoms in a supercell. 
We relax the structure with a quasi-Newton method in which the atomic forces are predicted by the machine learning model and compare the optimized structure with the DFT result. 
As shown in Fig.~\ref{f2} (b) and (c), the out-of-plane distortions induced by the Moir\'e potential given by the DFT calculation and machine-learning model prediction match very well. 
These results align with our expectations since the DPMD model has been proven successful and robust in numerous cases for atomic force prediction~\cite{Yang23p829,Niu20p2654,Rodriguez21p55367,Piaggi22pe2207294119,Chen23p053603,Liu22p024503,Wu23p144102,Wu21p024108,Wu21p174107}.

\subsection{Model performance for electronic band predictions}

\begin{figure*}
\centering
\includegraphics[width=16.0cm]{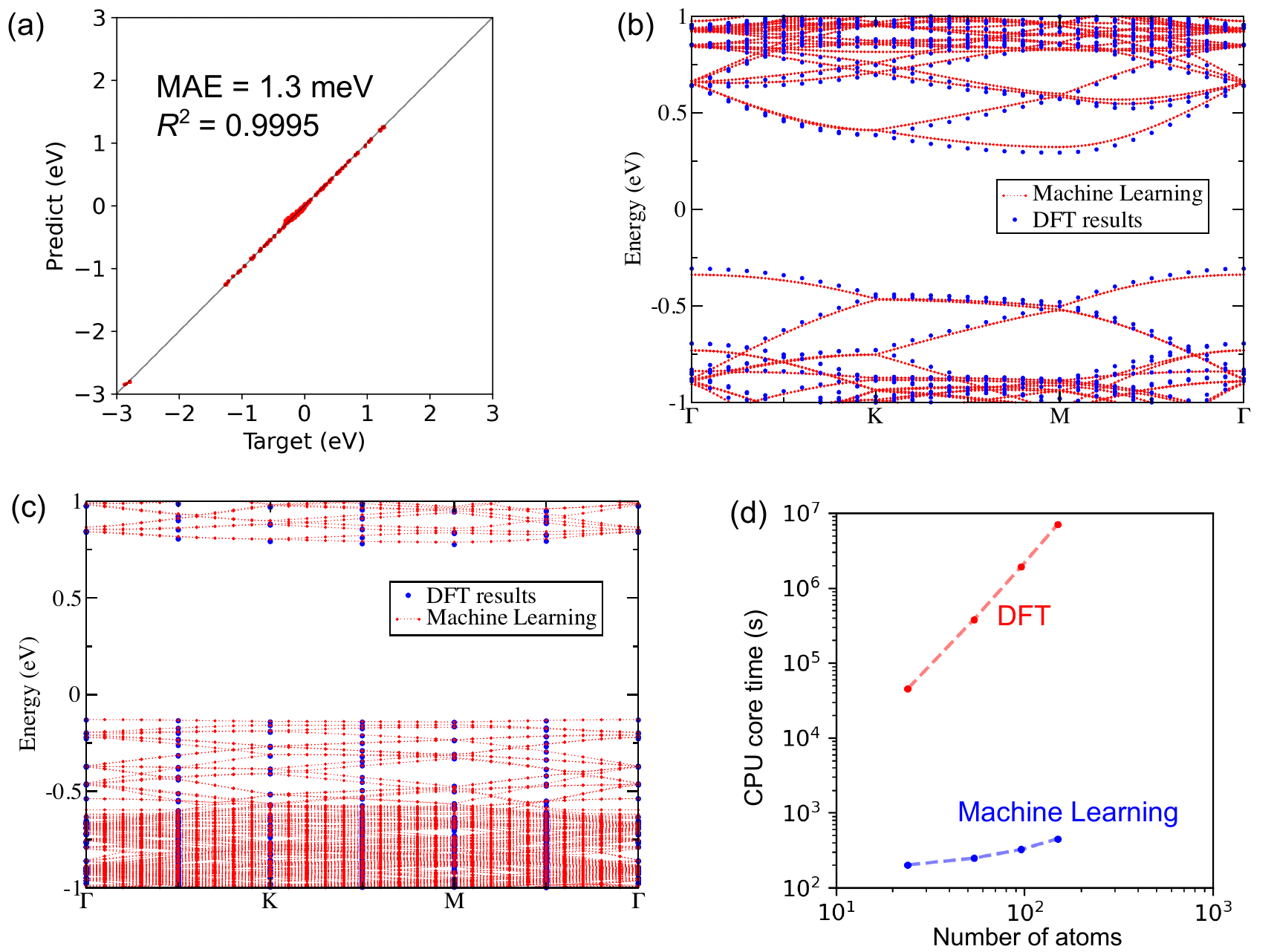}
\caption{\textbf{Performance of the WANDER for electronic band predictions. } \textbf{a,} The parity plot for Wannier Hamiltonian elements for the 8.95$^{\circ}$-twisted bilayer MoS$_2$. 
\textbf{b,c,} Comparisons between the electronic band structures given by the WANDER and DFT calculations for \textbf{(b)} an 8.95$^{\circ}$-twisted bilayer MoS$_2$ structure with 201 atoms and \textbf{(c)} a 3.42$^{\circ}$-twisted bilayer MoS$_2$ structure with 1308 atoms.  
\textbf{d,} Computational costs of DFT and WANDER for calculating the electronic bands of structures with different numbers of atoms.
}
\label{f3}
\end{figure*}

We use the WANDER to predict the Wannier Hamiltonian elements of the 8.95$^{\circ}$-twisted bilayer MoS$_2$, consisting of 201 atoms. 
We then compare these predictions with the results obtained from DFT calculations and Wannier90, illustrated in Fig.~\ref{f3} (a). 
The MAE is 1.3 meV, indicating an exceptionally high level of machine learning prediction accuracy. 
The electronic band structure, generated through Wannier interpolation using the predicted Wannier Hamiltonian, closely aligns with the DFT result, as depicted in Fig.~\ref{f3} (b).

To further validate our model, we extend its application to predict the band structure of a 3.42$^{\circ}$-twisted bilayer MoS$_2$ structure composed of 1308 atoms. 
This prediction is then compared with the results obtained from DFT calculations accelerated by GPUs. 
As illustrated in Fig.~\ref{f3} (c), our predicted results exhibit a strong agreement with the DFT calculations, providing additional evidence for the robustness of our model.
In Fig.~\ref{f3} (d), we present a comparative analysis of the efficiency between our model and DFT calculations [using the Generalized Gradient Approximation (GGA) functional]. 
Our model can generate electronic band structures with an accuracy comparable to DFT calculations but with an acceleration in computational speed of $10^3{\sim}10^4$ times.

\begin{figure*}
\centering
\includegraphics[width=17.0cm]{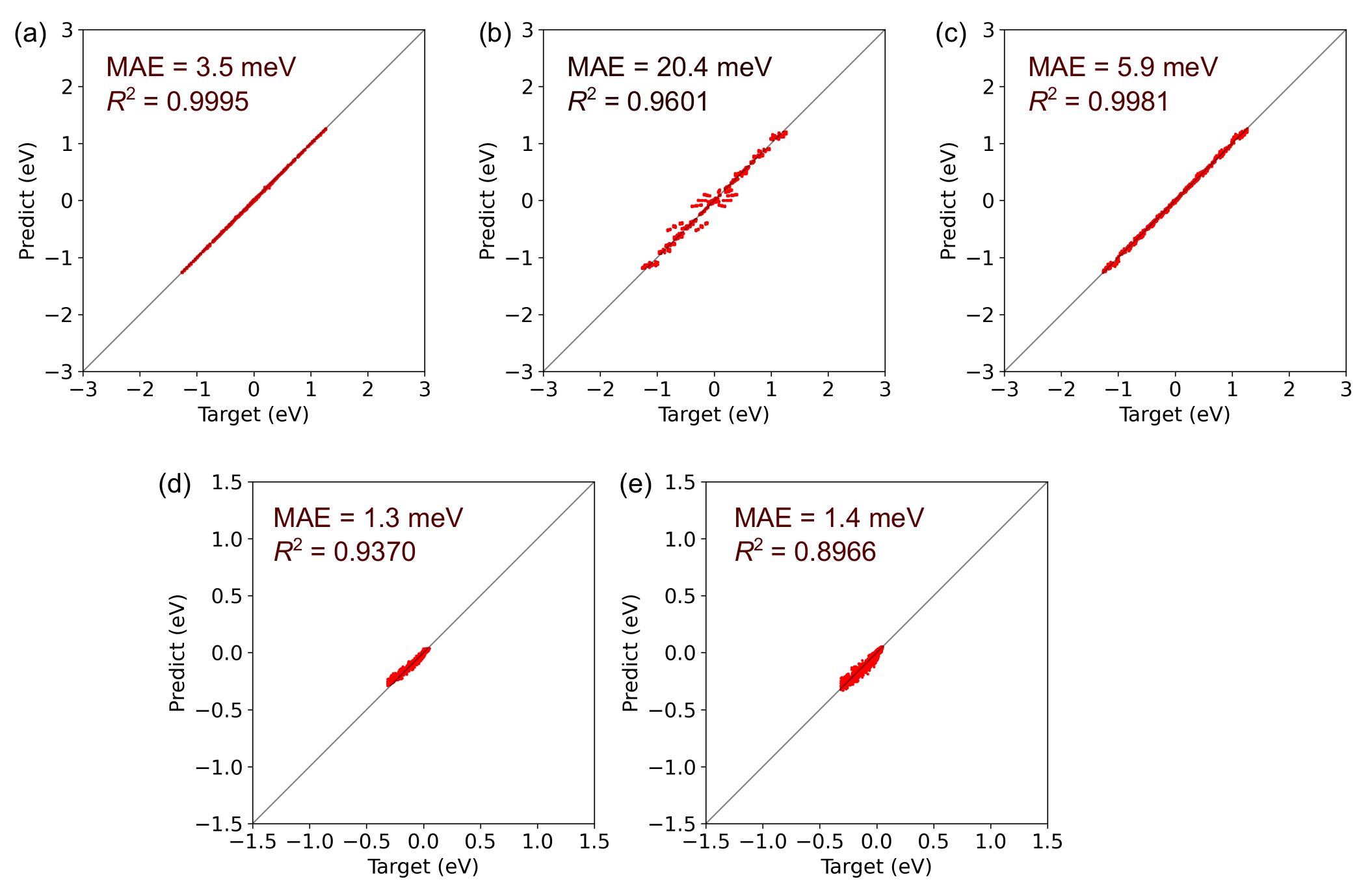}
\caption{\textbf{Comparisons of the performances of models with different inputs in predicting Hamiltonian elements} \textbf{a-c,} The parity plots for different models predicting intra-layer Hamiltonian elements of structures in the testing dataset. 
The inputs are \textbf{(a)} RAERs plus relative-position vectors (adopted in the model shown in Fig.~\ref{f1}), \textbf{(b)} relative-position vectors only, and \textbf{(c)} inputs in DPMD plus relative-position vectors.
\textbf{d,e} The parity plots for different models predicting inter-layer Hamiltonian elements of the 8.95$^{\circ}$-twisted bilayer MoS$_2$ structure with 201 atoms. 
Here, we focus on transferability so that we consider a twisted structure that is much different from the structures in the training dataset.
The inputs are \textbf{(d)} relative-position vectors only (adopted in the model shown in Fig.~\ref{f1}), and \textbf{(e)} RAERs plus relative-position vectors.
}
\label{f4}
\end{figure*}

\section{Discussion}

The selection of different inputs representing different categories of Wannier Hamilton elements is based on the underlying physical mechanisms and improves the performance of the model. 
The intra-layer hopping integrals primarily depend on the vector between two atoms. 
Most of the tight-binding models, which can simulate electronic band structures with reasonable accuracies, express the hopping coefficients as functions of inter-atomic distances only~\cite{Liu13p085433,Lenosky97p1528,Tang96p979,Ribeiro11p235312}. 
However, our work shows that the local atomic environments play a secondary but nonnegligible role. 
We consider a testing model in which only the relative-position vector is used as the input for predicting intra-layer hopping integrals.
The performance of this testing model [as shown in Fig.~\ref{f4} (b)] is noticeably worse than the WANDER [as shown in Fig.~\ref{f4} (a)], even though they share the same hyperparameters. 
These results demonstrate that information about local atomic environments serves as an essential input for a higher-accuracy model.

We consider another testing model in which the inputs for the two atoms in the DPMD model are stacked with their relative-position vector as the input for intra-layer hopping integral prediction. 
However, even with the same hyperparameters, the performance of this testing model is inferior to our WANDER based on RAER, as shown in Fig.~\ref{f4} (c). 
The rationale behind this lies in the reduced-dimensional nature of RAERs, which lowers the number of training parameters, facilitating the development of a model with a substantial weight on the relative position vector.

It is worth mentioning that the fundamental concept of RAER closely aligns with that of an autoencoder~\cite{Hinton06p504,Xie21p1}. 
In an autoencoder, an input (such as the local atomic environment) undergoes a transformation into a code, and the original input is reconstructed through a decoding function. 
In this study, the last hidden layer (referred to as RAER) in the DPMD model serves as the code containing essential information about the local atomic environment. 
The utilization of RAER in this work serves to simplify the model, thereby enhancing the likelihood of achieving a model that effectively captures the underlying physics and, consequently, achieves a higher accuracy.

In predicting inter-layer hopping integrals, we employ relative-position vectors as inputs. 
This choice is motivated by the absence of atoms in the inter-layer gap affecting the Hamiltonian, and the hopping integral mostly depends on the relative positions of the atomic pair~\cite{Marzari12p1419}. 
Introducing local atomic environment information into the input could complicate the model and hamper its transferability; generally speaking, models with fewer parameters often exhibit better transferability in deep learning~\cite{Zhang21p107,Moradi20p3947}.  
Twisted structures are not included in the training dataset due to their high computational costs. 
Consequently, the transferability of the model becomes crucial.
Fig.~\ref{f4} (d) and (e) illustrates a performance comparison on an 8.95$^{\circ}$-twisted bilayer MoS$_2$ structure between our WANDER and a testing model using RAERs plus relative-position vectors as inputs. 
The WANDER exhibits significantly higher accuracy, as evidenced by an R-squared value of 0.9370 compared to 0.8966 observed in the testing model.

By carefully choosing input parameters for distinct categories of Wannier Hamiltonian elements, the WANDER is able to simulate atomic and electronic structures simultaneously with accuracy comparable to that of DFT but at a significantly lower computational cost. This dual-functional model shows great promise for applications in exploring interactions between structures and electronic behaviors, including phenomena like electron-phonon coupling and structural change-induced metal-insulator transitions, especially in large-scale systems.

As a departure from the DeepH model~\cite{Li22p367,Li23p321,Gong23p2848}, which is a pioneering work in machine-learning electronic structure simulations, the WANDER offers an alternative for predicting the electronic band structures of scientifically significant systems with low periodicities, such as twisted bi-layer and multi-layer 2D materials.
Moreover, a key emphasis of this work is showcasing how existing force fields can be endowed with the capability for electronic structure simulations. 
We chose the DPMD model as an illustrative example of this successful integration. 
However, the fundamental principles of this method, which involve utilizing the local atomic environment for on-site interaction prediction, employing a reduced atomic environment representation along with a relative-position vector for intra-layer hopping prediction, and using a relative-position vector for inter-layer hopping prediction, are rooted in physical laws. These design principles are expected to be applicable to any machine-learning force field model.

This model presents opportunities for enhancement across several dimensions. 
While it exhibits satisfactory transferability in the context of twisted bi-layer structures, its performance on other unseen structures remains unpredictable, which is a common challenge in many machine-learning models. 
Addressing this issue can be approached through the following strategies. 
Firstly, we can enhance the model by simplifying it, which can be achieved by reducing the number of free parameters~\cite{Zhang21p107,Moradi20p3947}. 
In this study, we designate the last hidden layer in the DPMD model as the RAER. A more in-depth exploration, such as employing the autoencoder method~\cite{Hinton06p504}, can be conducted to determine the minimal dimensionality required for the RAER.
Furthermore, expanding the dataset to include a broader range of structure categories can mitigate the challenge of encountering unpredictable structures. 
This approach necessitates the development of high-throughput computational methods for generating large amounts and standardized Wannier Hamiltonians serving as the training dataset~\cite{Vitale20p66}.

This work also serves as inspiration for additional investigations in various directions.
For instance, this work adopts Wannier functions, widely utilized in solid-state physics, as the basis.
Theoretically speaking, any physical quantities calculable based on Wannier functions, including but not limited to spin Hall conductivity~\cite{Wang06p195118,Zhou19p060408}, transport properties~\cite{Pizzi14p422}, electron-phonon coupling~\cite{Giustino07p165108, Ponce16p116}, and shift current~\cite{Ibanez18p245143}, can be integrated into the predictive scope of this model.
This work aims to pave the path toward developing multifunctional machine-learning models for simulating a wide range of physical quantities.

\section*{Method}
To prepare the dataset, we carry out DFT calculations on 4$\times$4 bilayer MoS$_2$
supercells using the \textsc{Quantum--espresso} package~\cite{Giannozzi09p395502etalp}.
The functionals used are generalized gradient approximation~\cite{Perdew96p3865}
and a 4$\times$4$\times{1}$ Monkhorst--Pack $k$-point mesh is used to sample the Brillouin zone~\cite{Monkhorst76p5188}.
The kinetic energy cutoff for wavefunctions is 50 Ry. 
The Van der Waals interaction is simulated with the DFT-D method~\cite{Grimme06p1787}.
The Wannier Hamiltonians are calculated with the Wannier90 package~\cite{Mostofi14p2309,Pizzi20p165902}.

When generating the input data for the DPMD model, 
$R_C$ is selected as 5.2\ \AA.
After acquiring the DPMD model, we optimize the twisted structure with a quasi-Newton method, in which 
\begin{equation}
\bm{x}_{i,n+1}=\bm{x}_{i,n}+\alpha\bm{f}_{i,n}.
\end{equation}
Here, $\bm{x}_{i,n}$ is the position of atom $i$ at the $n$th step.
$\bm{f}_{i,n}$ is the force on atom $i$ at the $n$th step.
$\alpha$ is set as 0.529\ \AA$^2$/Ry.

The model is constructed using the PyTorch Python library~\cite{Paszke19}. 
Before progressing into the hidden layers, batch normalization is applied to the input to accelerate deep network training by reducing internal covariate shifts. 
When training the models to predict atomic forces and intra-layer hopping integrals, the two hidden layers consist of 64 and 32 nodes, respectively.
For the model aimed at predicting inter-layer hopping integrals, the two hidden layers comprise 32 and 16 nodes, respectively.
For predicting the Wannier Hamiltonian elements, different neural networks are trained for different kinds of atomic pairs (Mo-Mo, Mo-S, S-Mo, or S-S) and different inter-atomic distance ranges (long-range interaction for $R>6$ \AA\ and short-range interaction for $R<6$ \AA).
The Adam optimizer is used in this work.

\section*{Data availability}
The datasets for training the models in this work are available at the Digital Repository Service of Northeastern University (http://hdl.handle.net/2047/D20630194)

\section*{Code availability}
The code for this work is available on GitHub (http://github.com/yuboqiuab/multifunctional)

\section*{Acknowledgements}
Y.Q. and Q.Y. were supported by the U.S. Department of Energy, Office of Science, Basic Energy Sciences, under Award No. DE-SC0023664. W.G. was supported by the U.S. National Science Foundation under grant No. DMR-2323469. The research used resources of the National Energy Research Scientific Computing Center (NERSC), a U.S. Department of Energy Office of Science User Facility located at Lawrence Berkeley National Laboratory, operated under Contract No. DE-AC02-05CH11231 using NERSC award BES-ERCAP0029544. We thank Fei Xue for the valuable discussions.

\bibliography{ref}

\end{document}